\documentclass{iau} 
\usepackage{graphicx}

\title[X-Ray Properties of Black Widows] 
{X-Ray and Optical Properties of Black Widows and Redbacks}

\author[Mallory S.E. Roberts et al.]   
{Mallory S.E. Roberts$^{1,2}$, Hind Al Noori$^1$, Rodrigo A. Torres$^1$,
Maura A. McLaughlin$^{3}$, Peter A. Gentile$^{3}$, Jason W.T. Hessels$^{4}$, Scott M. Ransom$^{5}$, Paul S. Ray$^{6}$, Matthew Kerr$^{6}$ \and Rene P. Breton$^{7}$}

\affiliation{$^1$New York University Abu Dhabi, Saadiyat Island, Abu Dhabi, UAE {\tt malloryr@gmail.com}\\
$^2$Eureka Scientific, Inc. Oakland, CA USA, $^3$West Virginia University, Morgantown, WV, USA\\
$^4$University of Amsterdam, Netherlands, $^5$NRAO, Charlottesville, VA, USA\\
$^6$U.S. Naval Research Lab, Washington D.C., USA, $^7$University of Manchester, UK
}

\pubyear{2017}
\volume{337}  
\jname{Pulsar Astrophysics -­ The Next 50 Years}
\editors{P. Weltevrede, B.B.P. Perera, L. Levin Preston \& S. Sanidas, eds.}

\begin{document}

\maketitle

\begin{abstract}
Black widows and redbacks are binary systems consisting of a millisecond 
pulsar in a close binary with a companion having matter 
driven off of its surface by the pulsar wind. X-rays due to an intrabinary 
shock have been observed from many of these systems, as well as orbital
variations in the optical emission from the companion due to heating and 
tidal distortion. We have been systematically studying these systems in 
radio, optical and X-rays. Here we will present an overview of X-ray and 
optical studies of these systems, including new XMM-Newton and NuStar
data obtained from several of them, along with new optical photometry.

\keywords{pulsars: general, pulsars: individual (PSR J2129$-$0429), binaries: eclipsing}
\end{abstract}

\firstsection 
\section{Introduction}

Black widows and redbacks, collectively known as spiders, are millisecond pulsars (MSPs) in close ($P\lesssim 1$~day) orbits around stars which have or have had material driven off of them by the pulsar wind (cf. \cite[Roberts 2011]{rob11}). As in the case of the original Black Widow pulsar (\cite[Fruchter \etal\ 1990]{fbb+90}), this material often causes 
radio eclipses as well as X-ray emission through a shock interaction with the 
pulsar wind (\cite[Arons \& Tavani 1993]{ar93}). The shock can heat the pulsar facing surface of the tidally-locked companion leading to significant orbital modulation of the 
optical emission. The companion may also be nearly Roche lobe filling and so 
will be tidally distorted which can cause ellipsoidal  modulation as well (cf. \cite[Breton \etal\ 2013]{bre13}). 
Orbital period variations are often seen from spiders, with both positive and negative period derivatives sometimes even from the same source (cf. \cite[Arzoumanian \etal\ 1994]{aft94}). The magnitude of these changes is much larger than expected from gravitational radiation 
and may result from interaction between the pulsar wind and the companion's magnetic field via the \cite[Applegate \& Shaham (1994)]{as94} mechanism. 

Black widows and redbacks are differentiated by the mass of their companion. Theoretically, there is a tight correlation between the orbital period and evolutionary end state white dwarf mass for a system where the pulsar's spin-up to millisecond periods was driven by Roche-lobe overflow (\cite[Rappaport \etal\ 1995, Ergma \etal\ 1998, Tauris \& Savonije 1999]{rpj+95,esa98,ts99}). While most MSPs are consistent with this relationship, black widow companions have masses which are well below what is expected for a white dwarf and redbacks have masses which are above what is expected. In the former case, this may be due to there having been an extended period of ablation of the companion by the pulsar wind which may ultimately result in an isolated MSP (\cite[van den Heuvel \& van Paradijs 1988, Ruderman \etal\ 1989, Ergma \& Fedorova 1991]{vp88,rst89,ef91}). 
In the latter case, the companion has not yet finished its evolution and may be experiencing a short term interruption of the accretion process (\cite[Podsiadlowski \etal\ 2002, Benvenuto \etal\ 2014]{prp02,bdh14}). 
Several redback systems have now been observed to transition between ablating and accreting states on timescales of months to years (\cite[Archibald \etal\ 2009, Papitto \etal\ 2013, Bassa \etal\ 2014]{arc09,pfb+13,bph+14}), 
much shorter than what was predicted by evolutionary models. 

X-ray studies of these sources allow us to study pulsar winds much closer to the pulsar ($\sim 10^4$ light cylinder radii) than the termination shocks of pulsar wind nebulae like the Crab ($\sim 10^8$ light cylinder radii), and 
so may give insight into the magnetization properties of the wind near where it is launched (see \cite[Arons 2012]{aro12} for a review of pulsar winds and their shocks). 
Optical studies can tell us about the inclination of the system, the 
Roche-lobe filling fraction and heating of the companion, and potentially the 
masses of the components (\cite[van Kerkwijk \etal\ 2011, Breton \etal\ 2013]{vbk11,bre13}). 

The number of known, nearby spiders has increased greatly in the last 10 years 
due to targeted searches of $Fermi$ $\gamma$-ray sources (\cite[Hessels  \etal\ 2011, Ray \etal\ 2012]{hes11,rap+12}) and large scale pulsar surveys which routinely use acceleration searches to detect short period binaries such as the PALFA survey at Arecibo 
(\cite[Stovall \etal\ 2016]{sab+16}), the HTRU survey at Parkes (\cite[Bates \etal\ 2011]{bbb+11}), and the Drift Scan (\cite[Lynch\etal\ 2013]{lbs+13}) and GBNCC (\cite[Stovall \etal\ 2014]{slr+14}) surveys at Green Bank. Here we summarize some recent X-ray and optical studies of the newly discovered spiders. 

\begin{figure}[b]
 \vspace*{-0.75 cm}
\begin{center}
 \includegraphics[width=3.4in]{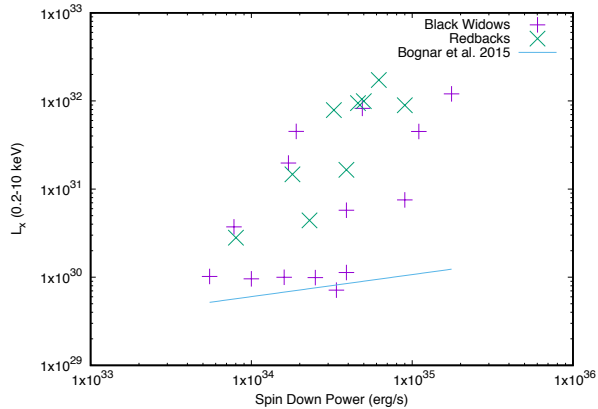} 
 \vspace*{-0.5 cm}
 \caption{X-ray luminosity $L_x$ vs. spin-down power $\dot E$ for spiders as observed by $Chandra$ and $XMM-Newton$. The line is the $L_X$ vs. $\dot E$ relationship of \cite[Bognar \etal\ (2015)]{brc15} for blackbody thermal emission observed from MSPs $L_{bb} = 10^{21.28}\dot E^{0.25}$erg/s.}
   \label{fig1}
\end{center}
\end{figure}

\section{X-Ray Observations of Spiders}

In \cite[Gentile \etal\ (2014) and Roberts \etal\ (2015)]{gen14,rmg+15}, we summarized observations of several 
black widows and redbacks with $Chandra$, noting that low count statistics 
often made difficult characterization of the X-ray orbital light curves. However, some general conclusions could be made. First, that many  black widows did not have much shock emission, but have largely thermal spectra similar to ordinary MSPs. Those black widows with significant non-thermal emission seemed to show a variety of orbital light curves, with only some, like the original Black Widow PSR B1957+20 (\cite[Huang \etal\ 2012]{hkt+12}), peaking near pulsar superior conjunction. 

On the other hand, redbacks seem to be, as a class, brighter in X-rays than black widows and show a more consistent light curve morphology, with a double peak around pulsar inferior conjunction. This may be related to the pulsar companions
subtending a much larger solid angle on the pulsar sky and a larger fraction of them being nearly Roche-lobe filling. The spectra tend to be very hard, with 
power law photon indices $\Gamma\sim 1.0-1.3$. This is harder 
than what is expected from ordinary shock acceleration (\cite[Sironi \etal\ 2015]{skl15}), and may be an indication of magnetic reconnection playing an important role (eg. \cite[Sironi \& Spitkovsky 2011]{ss11}). NuStar observations of the redbacks PSR J1023+0038 (\cite[Tendulkar \etal\ 2014]{tya+14}) and 
PSR J1723$-$2837 (\cite[Kong \etal\ 2017]{kht+17}) show this hard spectra can extend out to several tens of keV, 
and requires very efficient conversion of spin-down power into X-ray emission. We have obtained a NuStar observation of the redback PSR J2129$-$0429 which 
samples 2 orbits.  This also has a $\Gamma \sim 1.1-1.2$ spectrum extending to at least $\sim 30$~keV. However, these data show the second peak having some spectral variation. Full results of this observation will be presented by Al Noori \etal\ (in prep). 

The phase of the double peak in redbacks is somewhat of a puzzle. Models based 
on a wind-wind shock between the pulsar wind and that of the companion can 
reproduce this kind of light curve, but only if the companion wind momentum is many times that of the pulsar (\cite[Romani \& Sanchez 2016, Wadiasingh \etal\ 2017]{rs16,whv+17}). Since it is unreasonable to expect a $0.2-0.4~M_{\odot}$ 
star to have winds carrying $\sim 10^{35}$~erg/s of energy, quasi-Roche lobe overflow (\cite[Benvenuto \etal\ 2015]{bdh15}) has been invoked. However, this might not result in a very stable light curve. Archival $Swift$ data and the $NuStar$ observations suggest the double peaked curve of PSR J2129$-$0429 seen by $XMM-Newton$ is a consistent feature (Al Noori \etal\ in prep). It may be that a 
strong companion magnetic field plays an important role in shaping the light curve. 

In order to improve statistics on the X-ray properties of spiders, we obtained $XMM-Newton$ observations of 4 recently discovered black widows and 2 redbacks which had not been previously observed in X-rays over an entire orbit. We also obtained a full orbit on the redback PSR J2215+5135 which had been observed previously by $Chandra$ (\cite[Gentile \etal\ 2014]{gen14}). The overall flux from these sources and of previously observed spiders are summarized in 
Fig.\,\ref{fig1}. Around half of the black widows have emission which largely 
consists of thermal emission from the pulsar surface, with the rest having a clear non-thermal component. Even with $XMM-Newton$, it was difficult to obtain clearly defined orbital lightcurves from black widows. Redbacks, on the other hand, seem to nearly always 
have a significant non-thermal component. The $XMM-Newton$ lightcurve of PSR J2215+5135 clearly shows a double peaked lightcurve around pulsar inferior conjunction similar to what is seen in other redbacks (Gentile \etal\ in prep). 

\section{Optical Studies of Spiders}

Black widow companions seem to mostly be intrinsically cool and faint, with night side temperatures $T_{ns} \lesssim 4000$~K and optical magnitudes $m_V \gtrsim 23$ (cf. \cite[Breton \etal\ 2013]{bre13}). Their optical orbital lightcurves are completely dominated by heating of the companion by the pulsar with irradiation heating efficiencies on the order of 10\% - 30\% . Redbacks, on the other hand, have companions which are intrinsically very bright compared to the typical white dwarf companions of MSPs, with spectra and temperatures very similar to ordinary main sequence $g$ stars despite having masses of only $M_c \sim 0.15-0.5~M_{\odot}$. 
Las Cumbres Observatory (LCO) 1~meter observations of a number of the new black widows generally resulted in upper limits of optical emission with SDSS filter magnitudes $r^{\prime}$ and $i^{\prime} \gtrsim 21.5$. However, we have detected several redbacks and have been monitoring them. 

In the case of the redback PSR J2129$-$0429, \cite[Bellm \etal\ (2016)]{bkb+16} noticed a dimming trend in the $r^{\prime}$ band over the period between 2003 and 2015. We have been monitoring the source with the LCO since 2014 in the $g^{\prime}$, $r^{\prime}$ and $i^{\prime}$ filters and observed a continuation of this dimming trend until mid-2016 with color changes consistent with cooling. Since then, there seems to have been an increase in emission (Al Noori \etal\ in prep). We have also been monitoring PSR J2215+5135 since late 2014 and that system also seems to have brightened by $\sim 0.1$ magnitudes since then (Gentile \etal\ in prep). While redbacks have been observed to increase in optical brightness by $\sim 1$ magnitude when they transition to an accreting state (eg. \cite[Archibald \etal\ 2009]{arc09}), there has yet to be observed changes leading up to a transition. 

This work made use of data obtained from the Las Cumbres Observatory, $XMM-Newton$, $NuStar$, and $Swift$. Portions of this work at NRL were supported by the Chief of Naval Research.

\end{document}